\documentclass[a4paper,11pt]{article}
\pdfoutput=1

\usepackage{jinstpub}

\usepackage{lineno}

\usepackage{siunitx}

\graphicspath{{Figs/}}

\title{Techniques for mass production of large-sized GEM foil by the Korean CMS group for CMS phase-2 upgrade}

\author{I. Yoon, \note{Corresponding author.}}
\affiliation{Seoul National University,\\1 Gwanak-ro, Gwanak-gu, Seoul 08826, Republic of Korea}

\emailAdd{inseok.yoon@cern.ch}

\abstract{
  This study presents techniques for the mass production of large-sized GEM foils for the CMS phase-2 upgrade by the Korean CMS group. 
  The foil production facility is designed with a focus on mass production, including the adoption of the double-mask technique. 
  A polyimide wet etching technology that uses mono ethanolamine is reported, providing a safer working environment due to its lower inhalation toxicity compared to ethylene diamine. 
  The study also covers the denaturation of the etchant over time and the process of retuning. 
  Finally, R\&D results on soldering surface mount resistors with hot air for faster production are discussed.
}

\collaboration[c]{on behalf of CMS Muon Group}


\proceeding{The 7$^{th}$ International Conference on Micro Pattern Gaseous Detectors 2022\\
  December 11-16, 2022\\
  Weizmann Institute of Science, Rehovot, Israel}

\begin{document}
\maketitle
\flushbottom

\section{GEM foil production by the Korean CMS group}
\label{sec:overview}
Several upgrades of the CMS detector are ongoing for the luminosity upgrade of the Large Hadron Collider (HL-LHC).
One of these upgrades is the installation of three detector stations based on the Gas Electron Multiplier (GEM) technology, known as the ME0, GE1/1, and GE2/1 stations, in order of their distance from the interaction point.
The production and the installation of the GE1/1 system was completed in 2020.
Production and assembly of the GE2/1 detectors has begun and will be followed by the production of the ME0 detectors.

The Korean CMS group (KCMS), an association of Korean research institutes participating in the CMS experiment, has been designated as the second supplier of large-sized GEM foils together with the CERN Micro Pattern Technologies workshop (MPT) for the CMS GEM upgrades.
KCMS is contributing significantly to the upgrade by producing GEM foils equivalent to half of the GE2/1 and all of the ME0 stations.
So far, KCMS has produced 292 GE2/1 foils that have passed the quality control (QC) criteria established by the CMS GEM upgrade projects.

KCMS formed a consortium with Mecaro Co., Ltd. for the production of the foil, but after the consortium ended, the equipment for the production is being moved to the Institute of Basic Science (IBS - South Korea) for the production of ME0 foils.

\section{Production processes}
KCMS's GEM foil production facilities have been optimized for faster mass production.
In the next section, we will discuss the specific aspects of KCMS's GEM foil production technology that distinguish it from that of CERN MPT.

\subsection{Bipolar photolithograpy}
\begin{figure}[htbp]
	\centering
	\includegraphics[width=.25\textwidth]{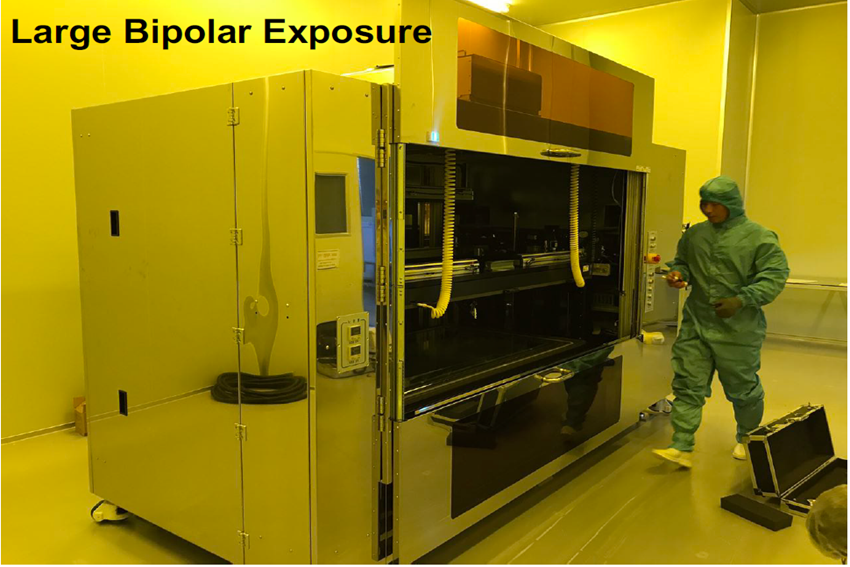}
	\qquad
	\includegraphics[width=.25\textwidth]{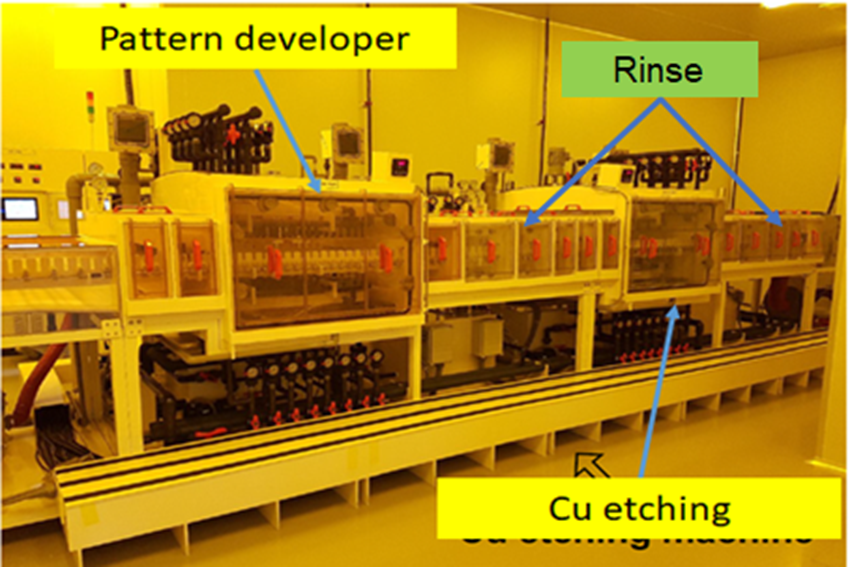}
	\caption{\label{fig:facilities} (Left) A large-sized bipolar UV exposure. (Right) A pattern developer and Cu etcher.}
\end{figure}

KCMS uses the double-mask technique for producing large-sized GEM foils, whereas CERN MPT is focusing the single-mask GEM foil production only \cite{Pinto}.
The definitions of the double and single-mask technique are described in \cite{Pinto}.
The double-mask technique simplifies the production process and enables faster production than the single-mask technique.
However, mask alignment becomes crucial.
The left side of figure~\ref{fig:facilities} shows a large-size bipolar UV exposure machine that not only transcribes patterns but also aligns the top and bottom sides of the masks.
The machine can align masks with a misalignment of less than \SI{5}{\micro\meter}, while the maximum allowed misalignment for optimal operation is \SI{7}{\micro\meter} (\SI{10}{\percent} of the hole diameter).
Emulsion glass masks are required for the alignment process.
The machine can form patterns of up to \SI{125}{\centi\meter}$\,\times\,$\SI{58}{\centi\meter} in size.

\subsection{Development and copper etching}
On the right side of figure~\ref{fig:facilities}, the pattern developer and copper (Cu) etcher machines are integrated, with a built-in conveyor belt to enable faster production.
When the flexible Cu clad laminate (FCCL) that has completed the photo process is inserted, it reacts with the reactant sprayed while moving on the conveyor belt.
The diameter of the developed hole can be adjusted by the moving speed of the belt, the concentration, and spray pressure of the reactant.

\subsection{Dry film resist stripping}
To strip the dry film resist, the processed FCCL is manually dipped into a bath filled with sodium hydroxide (NaOH).
NaOH is chosen for its fast reaction rate.

\subsection{Polyimide etching}
Polyimide (PI) etching is done by manually dipping the processed FCCL into a bath filled with an etchant mixture of potassium hydroxide (KOH) and monoethanolamine (MEA).
Like ethylenediamine (EDA) \cite{Pinto}, MEA exhibits anisotropic etching properties .
By adjusting the ratio of MEA and KOH, which is an isotropic etchant, the geometry of the developed hole can be tuned.
The geometry is not too sensitive to the duration of etching.
MEA is chosen for a safer working condition because the acute lethal concentration (LC) for mice exposed to MEA vapor is greater than \SI{2430}{\milli\gram\per\cubic\meter} \cite{PubChem_MEA}, while the median LC of EDA vapor is \SI{300}{\milli\gram\per\cubic\meter} \cite{PubChem_EDA}.
The etchant should be calibrated before use because the etchant exposed to air is slowly denatured by reacting with carbon dioxide.
The etchant can be calibrated through a few sample etchings.

\begin{figure}[htbp]
	\centering
	\includegraphics[width=.3\textwidth]{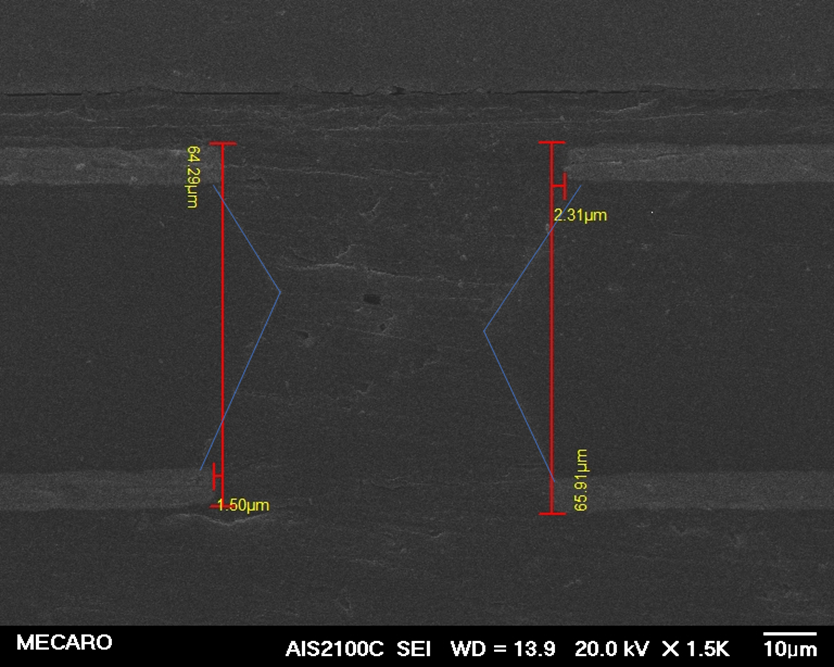}
	\caption{\label{fig:cross_section}
		A cross-sectional view of the produced GEM foil, where the blue lines were added by the author to enhance the visibility of the PI edge after taking the SEM image.}
\end{figure}

Figure~\ref{fig:cross_section} shows the cross-sectional view of the produced GEM foil with a PI hole of proper geometry, indicating acceptable mask alignment.
In this case, the residual misalignment is estimated to be at the level of \SI{3}{\micro\meter}, which satisfies the prescribed value of \SI{7}{\micro\meter}.

\section{Low-cost soldering technique for mass production}
GEM foils typically require surface mount resistors for high voltage protection.
However, soldering these small resistors onto FCCL can be challenging as they are easily pushed out by soldering tips, and excessive solder may reduce the gap between foils and cause sparks.

To address these challenges, KCMS has developed a low-cost soldering technique using a cream solder dispenser and hot air rework station.
The air pressure-operated solder dispenser deploys a uniform and accurate amount of cream solder, and a jig helps align the resistors and prevent GEM foils from sagging during the soldering process.
With this technique, the soldering process can be completed more quickly and accurately with less work fatigue.


\section{Validation of the production techniques}
The validation of the production techniques was conducted by assembling GEM detectors with the foils produced by the new vendor and measuring characteristics such as gain, gain uniformity, rate capability, aging property and other performance metrics \cite{ICHEP18_Validation}.
In addition, the techniques were verified again through the results of the
mass production for the CMS upgrades, described in section~\ref{sec:overview}.

\acknowledgments{
	We would like to express our gratitude to Dr. Rui De Oliveira (CERN) for his invaluable assistance.
	We also extend our thanks to Taeseong Jeong and Inseung Jeong (Mecaro) for their support.
	This study was supported by the National Research Foundation of Korea
}

\end{document}